# Selective detection of bacterial layers with terahertz plasmonic antennas

**Audrey Berrier**[*,1,†], **Martijn C. Schaafsma**[1], **Guillaume Nonglaton**[2], **Jonas Bergquist**[3], **Jaime Gómez Rivas**[1,4]

[1] FOM Institute AMOLF, Centre for Nanophotonics, c/o Philips Research Laboratories, HTC4, 5656AE Eindhoven, The Netherlands.
[2] CEA Leti, MINATEC Campus, Department of microtechnologiesfor Biology and Healthcare, 17 rue des martyrs, 38054 Grenoble, France.
[3] Analytical Chemistry, Dept. of Chemistry, Biomedical center and Science for life Laboratory, Uppsala University, PO Box 599, SE-75124 Uppsala Sweden
[4] COBRA Research Institute, Eindhoven University of Technology, P.O. Box 513, 5600 MB Eindhoven, The Netherlands .

[*]audrey.berrier@pi1.physik.uni-stuttgart.de
† now with Universität Stuttgart,Physikalisches Institut, Pffafenwaldring 57, 70550 Stuttgart, Germany

**Abstract:** Current detection and identification of micro-organisms is based on either rather unspecific rapid microscopy or on more accurate complex, time-consuming procedures. In a medical context, the determination of the bacteria Gram type is of significant interest. The diagnostic of microbial infection often requires the identification of the microbiological agent responsible for the infection, or at least the identification of its family (Gram type), in a matter of minutes. In this work, we propose to use terahertz frequency range antennas for the enhanced *selective* detection of bacteria types. Several microorganisms are investigated by terahertz time-domain spectroscopy: a fast, contactless and damage-free investigation method to gain information on the presence and the nature of the microorganisms. We demonstrate that plasmonic antennas enhance the detection sensitivity for bacterial layers and allow the selective recognition of the Gram type of the bacteria.

## 1. Introduction

Fast diagnosis of bacterial infections remains a major challenge in nowadays medicine. Despite the availability of antibiotics, bacterial infections are the cause of significant mortality around the world. The primary cause of mortality in this case is the late identification of the origin of an infection, which lead to inappropriate medical treatment. The overuse of large-scope, generic antibiotics induce over exposure of the environment, hence leading to the development of survival strategies from bacteria. As a result multi-resistant species are becoming more present, which causes real public health concerns. Therefore, identification of the nature of the infectious agents within a few minutes would represent a significant advance.

Most bacteria owe their mechanical stability to their outer cell wall. Its nature and thickness are the main distinctive features between the two main bacteria types: Gram positive or negative. A Gram-positive bacterium has a thicker cell wall than a Gram-negative bacterium. The Gram negative cell wall is chemically more complex and consists of at least two chemically distinct layers, whereas the Gram positive cell wall consists primarily of a polysaccharide component called peptidoglycan [1], [2]. Knowledge about the Gram type of bacteria is particularly useful since it allows direct use of the appropriate family of antibiotics and reduces the usage of generic broad screening range medicines [3].

Gram type determination is conventionally performed in a microbiological laboratory, and it takes a few days to get interpretable results. Other techniques like PCR (polymerase chain reaction) are very accurate but require expensive equipment, trained personal and take a few hours before giving results [4,5]. There is still a need for a fast and portable bacterial recognition device for point-of-care use.

Novel methods have been recently developed to detect the presence of a certain type of micro-organisms. Most of them make use of a pre-functionalization to capture specific bacteria [6,7] and optical methods operating in the visible and in particular based on evanescent waves (surface plasmon polaritons and waveguides). These techniques are not optimally adapted to interact with a monolayer of bacteria due to the size mismatch between the field decay length (some tens of nm) for the visible frequency range and the size of a bacterium (order of magnitude 1 μm). The use of long-wavelength radiation (i.e. far-infrared region) is therefore advantageous. The terahertz (THz) regime corresponds to the region of the electromagnetic spectrum between 0.1 and few THz. The wavelength of the radiation at 1 THz is 300 μm. Detection schemes of bacteria in the THz regime have been reported [8], although no selectivity was demonstrated. So far, few reports were concerned with the modification of the dielectric properties of bacteria as a function of their Gram type in the microwave region [9].

In this work, we present a method for label-free, functionalization-free, *selective and enhanced* detection of monolayers of different bacteria using THz radiation. The selectivity of the proposed method comes from the different dielectric response of biological matter at THz frequencies due to differences in their structural and chemical environment properties. The enhanced sensitivity is provided by THz plasmonics antennas. Plasmonics exploits the electromagnetic field enhancements originating from the collective oscillations of free electrons in a metallic material. In the particular case of a 3D structure, the electromagnetic field is confined to the vicinity of the structure, which supports localized surface plasmon polaritons (LSPPs). An example of such a 3D structure is a bowtie antenna, where two triangles with facing apices are separated by a small gap. When resonant incident radiation polarized along the longitudinal axis of the antenna interacts with the antenna, a peak in the extinction spectrum is observed, as well as a local field enhancement in its gap [10]. Such field enhancements in highly subwavelength areas are particularly interesting for sensing and spectroscopy of low volumes, where the interaction between the radiation and the substance to detect is greatly enhanced. At THz frequencies, semiconductor materials have a much lower permittivity than metals, which can moreover be tuned by varying the doping concentration [11]. It is advantageous to use semiconductor materials to form bowtie antennas in the THz range since the local field enhancements combine the plasmonic resonance with the lighting rod effect [10]. At THz frequencies the field decay of the evanescent tail of a LSPP is typically of a few microns. This localized field was used in a recent report to enhance the THz detection sensitivity of thin inorganic films [12]. THz plasmonics is hence a powerful candidate for enhanced bacterial layer detection. The combination of THz plasmonic structures with fully integrated, CMOS compatible, all-electronics THz sources and detectors could lead to ultra-compact, portable THz spectrometers used in practitioners' offices and point-of-care terminals [13].

**2. Terahertz spectroscopy and antennas**

Semiconductor plasmonic antennas operating in the THz range have recently been presented as novel plasmonic structures offering the possibility to actively control their THz response within a picosecond timescale [14]. Owing to the long wavelength of THz radiation, doped-silicon plasmonics antennas can be designed to interact optimally with a thin film [12]. We have fabricated silicon-based THz bowtie antennas following the process steps described in Ref [14]. We have started from commercially available quartz (fused silica) 6" wafers, and implanted Silicon-on-Insulator (SOI) wafers. The two wafers were bound together using a benzocylobutene (BCB) based process. After wet etching of the silicon substrate and the silica buffering layer, a 1.5 μm thick doped silicon layer remained on top of the quartz wafer. Bowtie antenna patterns were subsequently defined by conventional optical lithography and dry etching using the photoresist as etch mask. An example of *plasmonic chip* with quartz substrate of 1x1 cm$^2$ onto which a collection of silicon bowtie antennas have been etched, is shown in Figure 1. The bacterial layers were deposited on top of the plasmonic chips.

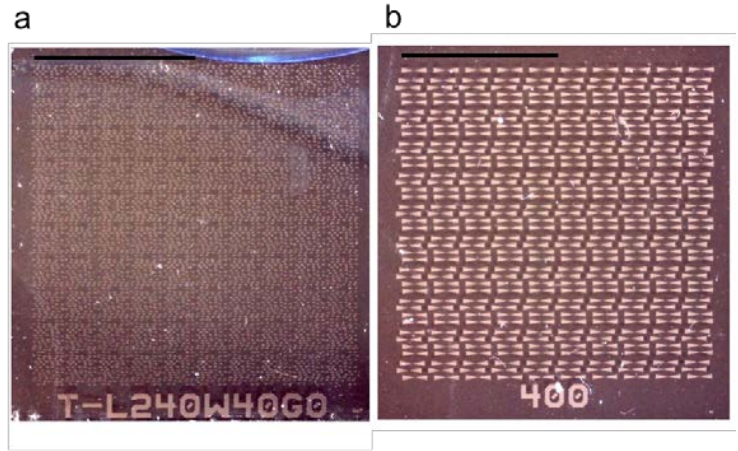

Figure 1 Photograph of the two plasmonic chips used in this work, a) high frequency design; b) low frequency design. The scale bars on top of the pictures represent 5 mm.

The resonant behavior of the plasmonic antennas depends on the material permittivity, i.e., on the doping concentration of the silicon, and also on the geometry of the antennas. In this work, the length of one of the triangles forming the bowtie antenna varies from 100 to 400 μm. By modifying the size and shape of THz plasmonic antennas, they can be designed to resonate in different frequency ranges. In particular, several designs of antennas are necessary to investigate the THz interaction with biological layers over the wide range 0.1 THz to 1 THz. In order to access both the high (0.5-1 THz) and low frequency (0.1-0.5 THz) regime, we have designed two types of antennas with resonances around 0.8 and 0.24 THz, respectively. Results obtained with both designs are reported in section 5.

Terahertz time-domain spectroscopy (THz-TDS) consists in measuring the transmission of single-cycle, picosecond THz pulses through a sample under investigation. This technique measures the electric field amplitude as function of time, hence providing information both on the amplitude and the phase of the electromagnetic field, and allows for picosecond time-resolution of the temporal response. The time-resolved terahertz time-domain spectrometer used in this work is based on a Ti:sapphire laser (Femtolasers, Fusion) with 75 MHz repetition rate. The output pulses at around 800 nm wavelength have a duration of about 20 fs. The output beam is split in two parts to generate and detect the THz probe pulse. The THz pulse is generated via a photoconductive antenna (Gigaoptics). The THz field is linearly polarized. The full width at half maximum of the THz beam was determined to be 1.7 mm using a knife-edge technique. The transmission of the THz pulse through the sample is measured in the detection path via electro-optic sampling in a ZnTe crystal followed by a quarter wave plate, a Wollaston prism and a pair of balanced silicon photodetectors. THz transients are recorded point by point using a delay stage between the THz beam and the sampling beam. Transmission spectra are obtained through Fourier transformation of the THz transient. We define the extinction $E$ as $E=1-T/T_{norm}$, where $T$ is the transmission (amplitude) through the plasmonic chip and $T_{norm}$ is the transmission through a bare quartz substrate. The resonances of the antennas of the plasmonic chip are characterized by a peak in the extinction spectrum.

### 3. Terahertz and bacteria

Terahertz spectroscopy is a technique allowing for the detection and identification of substances based on their chemical composition and in particular on their low energy vibration modes. THz radiation probes collective structural vibrational modes, as well as the network of hydrogen bonds. Water and other polar liquids absorb strongly THz radiation. In general, biological material can be described by a resonant and a relaxational response [15]. The relaxational response is usually dominant and gives rise to a smooth spectral variation of the permittivity. The different chemical composition, but also the intermolecular character of the molecular vibrations/rotations, leads to differences in the permittivity in the THz regime. Therefore, it is expected that Gram positive and Gram negative bacteria, having different chemical nature, will interact differently with THz radiation.

The use of plasmonics is a powerful strategy to enhance the interaction of the probing field with the matter to be detected. At optical frequencies, surface plasmon resonance (SPR) sensors make use of the field enhancement at the interface of a metal and its dielectric environment to give rise to highly sensitive sensors. However, SPR sensors operating in the visible are very sensitive only for the detection of very thin layers (a few 10's of nm) [16]. Therefore this technique is not the most efficient for large entities such as cells with typical sizes of microns.

The advantage of the THz regime for the detection of cells is the long wavelength of the THz radiation. It is illustrative to compare the thickness of a bacterial cell with the field profile at the surface of a silicon bowtie antenna.. As it is discussed in Ref. [12], this field decays typically over a distance of few microns and the overlap of the field enhancement profile with a bacterial monolayer is optimal. Therefore the overlap between the microorganisms and the THz radiation can be maximized and the detection sensitivity optimized.

**4. Bacterial layers**

*4.1 Description and preparation of the bacterial layers*

The bacteria types used in this study are *E.coli* strain *DH5a (ATCC PTA-3137)*, *B.subtilis* strain *168 (ATCC 23857)*, *S.marcescens (ATCC 8100)*, *S.epidermidis (ATCC 14990)*, and *M.catarrhalis (ATCC 25240)*. Optical microscope pictures of deposited cells on a quartz substrate are displayed in Figure 2. *E.coli* and *M.catarrhalis* are Gram-negative bacteria, whereas *B.subtilis*, *S.marcescens* and *S.epidermidis* are Gram-positive types. *E.coli* is a rod-shaped bacterium commonly found in the lower intestine of humans and animals. While most *E.coli* strains are harmless, some strains – in particular those of fecal origin, can cause severe food poisoning by contamination of food and water. *B.subtilis* is a rod-shaped bacterium commonly found in soil and can cause food poisoning. *S.marcescens* is a rod-shaped bacterium which can cause nosocomial infections of urinary tract, wounds and gastrointestinal systems. *S.epidermidis* is a coccus-shaped bacterium which, although part of the human flora, can cause nosocomial infections in immuno-compromised patients. *M. catarrhalis* is a diplo-coccus (i.e., two joint rounded cells) shaped bacterium commonly involved in infections of the respiratory system, nervous system and joints of humans. All the five types of bacteria can be handled safely in a MLII class microbiological laboratory.

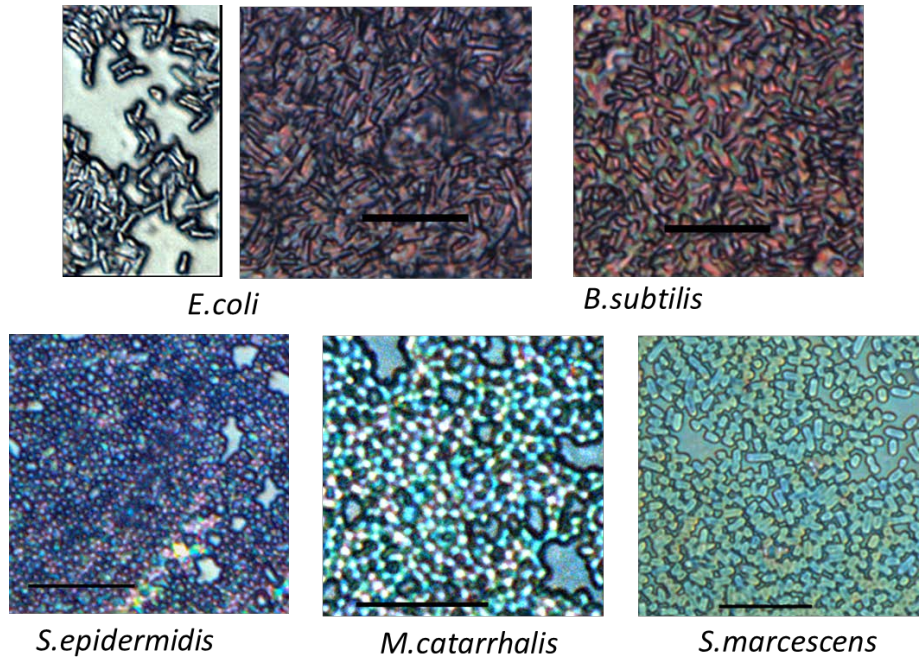

Figure 2 Optical microscope images of layers of bacteria deposited on a quartz substrate for the five species studied in this work: *E.coli*, *B.subtilis*, *S.epidermidis*, *M.catarrhalis* and *S.marcescens*. In all the pictures the scale bar represents 10 μm.

Bacterial stocks were bought from ATCC and glycerol stocks were prepared for storage at -80°C. The day before THz measurements were performed, a frozen sample of the bacterial stock was used to seed a culture using 10 ml of tryptone soy broth as culture medium. After overnight growth at 37°C, 1 ml of the culture medium was transferred to an Eppendorf tube and centrifuged at 3000 rpm for 5 min in order to extract the bacteria from the culture medium. The culture medium was subsequently replaced by a buffer solution. Two buffer solutions were used, the common phosphate buffer PBS and an isotone aqueous solution containing 1.2 g/L of ammonium acetate and adjusted at pH 7.

The bacterial layers were formed by depositing 12x5 μl of the bacterial solution onto an area of 1 cm$^2$. The excess water was dried on a hotplate set at the low temperature of 42°C in order not to affect the bacteria. The resulting surfacic concentration of bacteria is 0.8 cfu/μm$^2$ (with "cfu" standing for colony forming unit, i.e. the number of viable bacteria). Double layers were made by repeating the deposition procedure anew, and have hence a surfacic concentration of 1.6 cfu/μm$^2$.

We underline that the work reported in this section involves the handling of living microorganisms, as well as the incubation and growth of fresh bacterial solutions. We do not work with fragments of bacteria since we want to investigate the response of living bacteria to THz waves. THz transmission measurements were performed shortly after deposition (unless otherwise mentioned) in order to control the quality of the bacterial layers.

*4.2 Role of the buffer solution*

The homogeneity of the dried bacterial layer plays an important role in the outcome of the measurements. In particular, it was noticed that the buffer used to dilute the bacteria was of importance. A commonly used buffer solution is PBS (phosphate buffered saline solution). However PBS is not an appropriate buffer solution when the biological layers need to be dried. Indeed, when drying PBS gives rise to crystallization of the salt contained in the solution. The crystallization of these salts creates regular patterns on the substrate, as can be seen in Figure 3. These regular patterns with a periodicity on the order of 100 μm introduce unwanted scattering of the THz waves and therefore must be avoided. Trustable THz measurements are therefore difficult to obtain using this buffer. Efforts have been dedicated to find optimal conditions for the formation of homogenous bacterial layers. Homogeneous bacterial layers are obtained if an ammonium acetate solution is used instead of PBS (see Figure 3b). In particular it was found that the use of an ammonium acetate buffer prevented the formation of the crystallization patterns observed in the case of the PBS buffer.

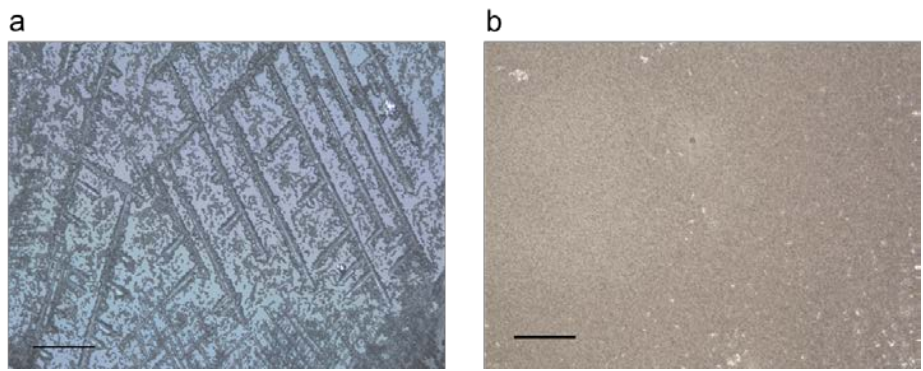

Figure 3 Optical microscope images of the dried bacterial layers a) using the phosphate saline buffer PBS, the scale bar indicates 100 μm ; b) using a ammonium acetate buffer, the scale bar indicates 200 μm.

*4.3 Thickness of the bacterial layer and deposition on THz plasmonic antennas*

It is important to know the thickness of the deposited bacterial layers. In order to evaluate it, we have deposited bacterial layers of several bacteria types on quartz substrates partially covered with thin strips of tape. The deposition protocol is identical to that used for antenna samples. Once the bacterial layer was deposited on the substrate, the stripe of tape was removed. Figure 4 displays a photograph of the samples after removal of the tape in the case of two different bacteria types, *E. coli* and *S. epidermidis*. Profilometer measurements were subsequently performed on the samples using the area without bacteria as a zero reference. Figure 4b displays the result of the profilometer measurements. The average thickness of the bacterial layer indicated by this measurement is between 500 nm and 1 μm, which corresponds to a monolayer of bacteria. Different bacterial layers have comparable thicknesses.

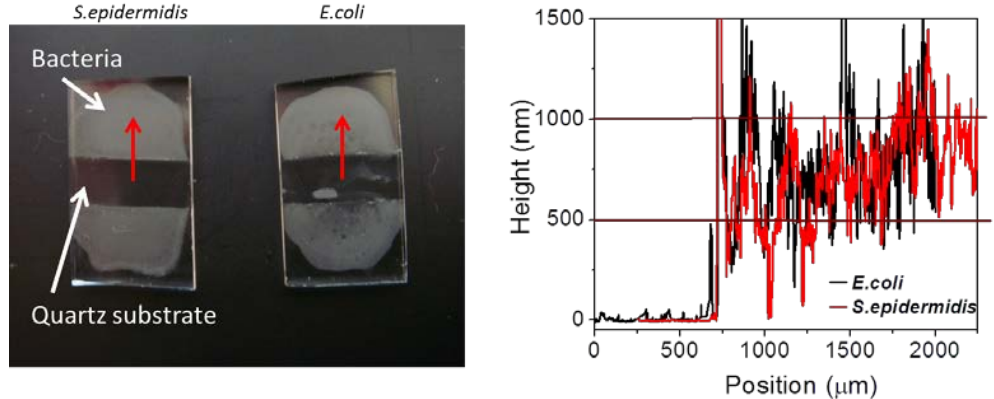

Figure 4 a) Photograph of the quartz samples with a deposited bacterial layer; b) Profilometer trace of the deposited bacterial layers. The arrows indicate the path of the profilometer on the samples.

The THz measurements are performed in transmission through the plasmonic chip. When deposited on top of the plasmonic antennas the coverage of the bacterial layer is conformal and the layers homogeneous, as can be seen in Figure 5. From this we can deduce that the thickness measured on a quartz substrate is also valid in the case of the deposition on the plasmonic antennas.

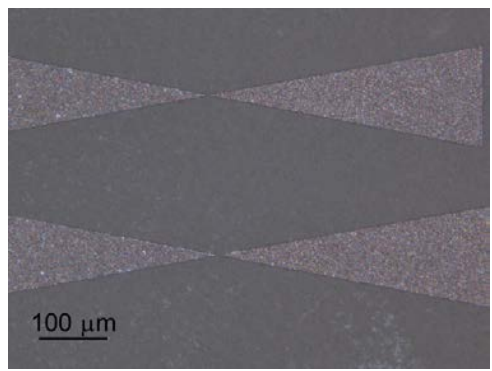

Figure 5 Optical microscope image of THz antennas covered by a monolayer of bacteria.

## 5. Results and discussion

### 5.1 Thin film detection of bacterial layers

We compare here the detection capability of the plasmonic chip to that of a quartz substrate. Subsequently, we have deposited a bacterial layer on the plasmonic chip and on the quartz substrate, and we have measured the THz transmission through both samples (as it is schematically represented in Figure 6(a)). The THz transmission through a quartz substrate without bacterial layer is used as a reference for both samples.

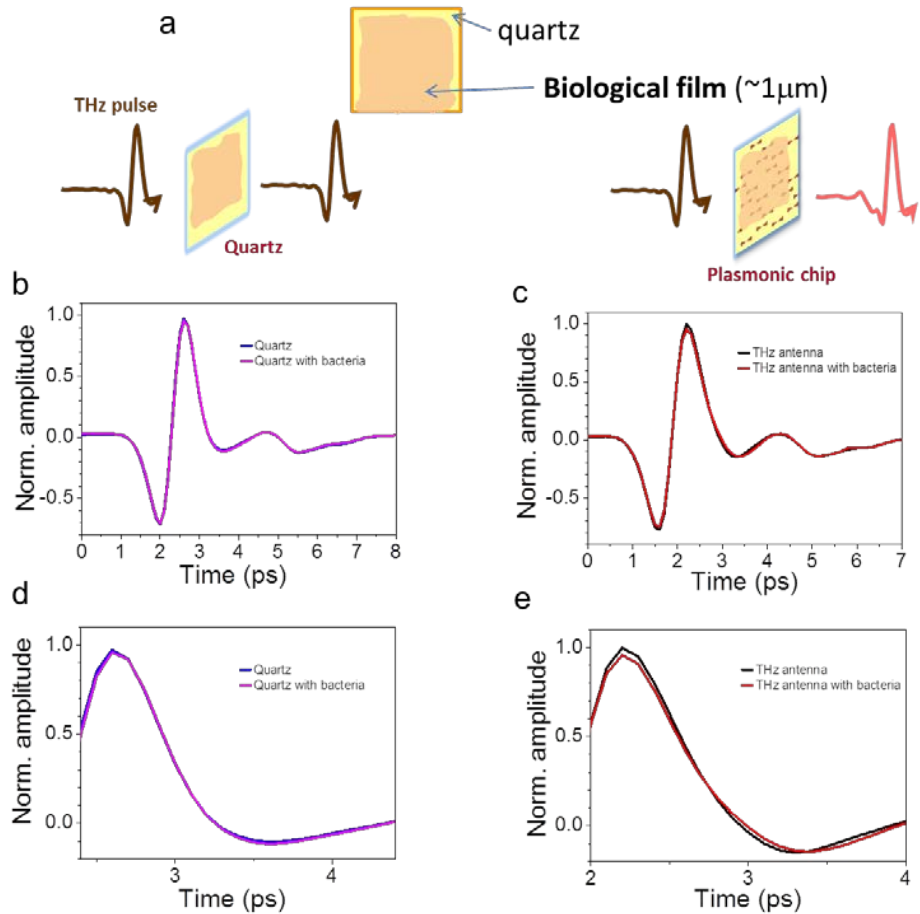

Figure 6 a) Schematic drawing of the measurement scheme. b) Time-resolved amplitude transient of transmission through a bare quartz substrate without and with a monolayer of Gram negative bacteria; d) is a magnified view of b). c) THz transient transmission through a sample with THz antennas with and without a monolayer of Gram negative bacteria. e) is a magnified view of c).

Figures 6 (b) and (d), where (d) is a magnified view of (b), display the THz transmitted transients through quartz substrates with and without a monolayer of bacteria. These measurements show that the bacterial layer is not detectable from the signal of the plain quartz substrate. To be detectable, the bacterial layer thickness should be of the order of the wavelength of the THz radiation, which is a few 100's of μm. As seen on Figure 4, the thickness of a monolayer of bacteria is around 1 μm. On the contrary, when bacterial monolayers are deposited on top of the THz antennas a measurable change in the THz transient is observed (Figure 6(c) and (e)). As discussed in section 2, these changes in the THz transients translate into a modification of the transmission spectrum that can be obtained by Fourier transformation of the time-domain signals. An example of Fourier transformed signal can be seen in Figure 7, where the black curve represents the extinction of the array of bare antennas and the red curve is the extinction of the array with a bacterial layer. The pronounced peak of extinction corresponds to the resonant extinction due to the excitation of localized surface plasmon polaritons in the individual bowtie antennas. The presence of the bacterial layer induces a shift of the resonance frequency due to the modification of the permittivity of the medium surrounding the antenna. Therefore, it is possible to detect the presence of the bacterial layer using the plasmonic chip. This represents a significant enhancement of the detection sensitivity for THz sensing of bacterial layers. In order to further increase the interaction of bacterial layers with THz radiation mediated via the plasmonic antennas, we have deposited double layers of bacteria. Figure 8 displays the change in the THz transient when the thickness of the deposited layer is doubled (thickness around 1.5 μm). The double bacterial layer gives rise, as expected, to a larger modification in the THz transient.

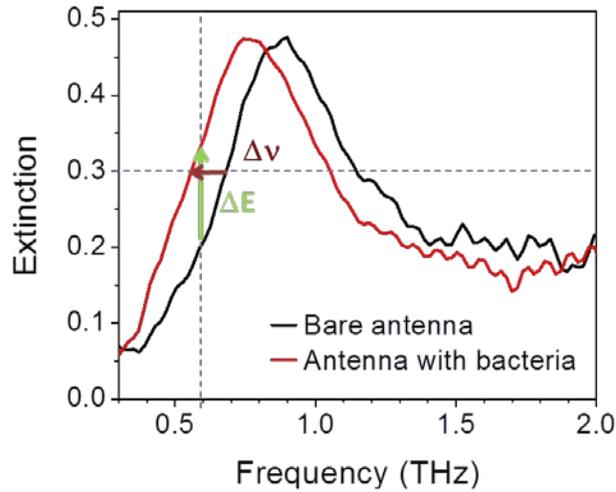

Figure 7 Example of Fourier transformed signal for a bare antenna sample and a sample with a bacterial layer. This figure also illustrates the determination procedure for the change in extinction ΔE and the frequency shift Δν by comparing the extinction values of the bare samples with samples with bacterial layer.

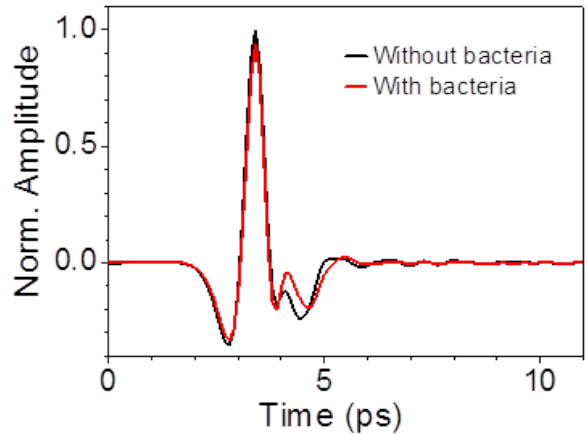

Figure 8 Comparison of the temporal THz transmission transients through a plasmonic chip with bare antennas and with a double bacterial layer on top.

It can be concluded that the presence of the THz antennas allows the detection of thin bacteria layers, down to the monolayer limit. In order to decrease the total number of bacteria in the detection area, there is a need for localized trapping of the bacteria in a limited area around the regions of high field enhancement of the antennas. Such localization can be brought with a versatile process using polyethylene imine combined to conventional photolithography. Results on this process are out of the scope of the present article and are reported in Ref. [17]. To reduce the detection sensitivity even further it is possible to measure a single THz antenna, instead of a collection of antennas as in the present work, by means of a conically tapered THz beam concentrator [18].

*5.2 Selectivity*

It is now interesting to look at the response of the THz plasmonic antennas to deposited layers of different bacterial types. The question of the selectivity of the detection mechanism is of major interest for sensors. Here we focus our attention on the determination of the Gram type of the bacterial layers. First, we start by depositing double layers of bacteria, for which the THz response is larger. We compare the spectral response of the antennas when covered by three different types of bacteria: *E.coli*, *B.subtilis* and *S.epidermidis*. Figure 9 indicates the spectral response of plasmonic antennas to double layers of bacteria. The extinction spectra are the combination of a resonant contribution (around the peak of the antenna resonance) and of a non-resonant contribution (outside of the resonance peak). The larger extinction response over the whole frequency range to the *E.coli* layer indicates that *E.coli* bacteria layers have a high THz absorption. This is clearly visible in the non-resonant part of the spectrum, but also in the broadening of the plasmonic resonance, as well as red shift of the resonance peak. It is remarkable to see that the two types of bacteria of Gram positive type (*B.subtilis* and *S.epidermidis*) behave in a very similar way and present a much lower absorption than the Gram negative type *E.coli*. The antenna resonance is in the latter cases red shifted but the peak broadening is much smaller, as well as the non-resonant absorption. This result suggests that the two bacterial types Gram positive and Gram negative have a different interaction with the THz waves.

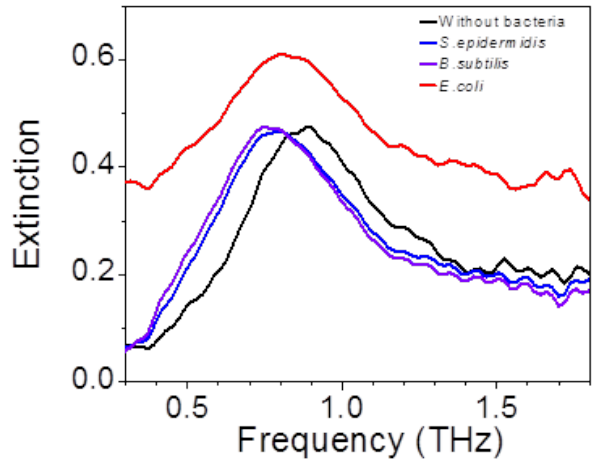

Figure 9 Extinction spectra of a plasmonics chip without bacterial layer and with a layer of *E.coli*, *B.subtilis* or *S.epidermidis*.

We now turn our attention to monolayers of bacteria to confirm the trend that Gram negative bacteria, like *E. coli*, present larger THz absorption than Gram positive bacteria, like *B. subtilis*. In order to reduce the influence of external perturbations, like changes in temperature, humidity or stability in THz setup, we have performed a large number of individual measurements (>50) spread over several measurement sessions on different days and we have analyzed the changes in shift of the antenna resonance due to the presence of the bacterial layer. As a purpose of comparison we define two quantitative parameters reflecting the increase in extinction at 0.6 THz, ΔE, and the red frequency shift for an extinction value of 0.3, Δν. The determination of those parameters is illustrated in Figure 7.

The relevant quantities are the difference of the values of ΔE and Δν obtained for the Gram positive and Gram negative bacterial layers during the same measurement session. We will refer to this difference as the differential signal and it is plotted in Figure 10 for the various measurement sessions. In this figure we plot the differential resonance frequency and the differential extinction for the antennas resonant at 800 GHz. The results show a positive differential signal for most of the measurement sessions. This means that a layer of Gram negative bacteria gives on average a larger extinction and a larger frequency shift than a layer of Gram positive bacteria. Measurements performed on antennas resonant at 240 GHz are reported in Figure 11. Here the differential extinction at 250 GHz, close to the antenna resonance, is presented. From the multiple measurements presented in this article, it can be stated that both at high and at low frequencies Gram positive bacteria (*B. subtilis* and *S. epidermidis*) can be distinguished from Gram negative bacteria (*E. coli*, *S. marcescens*, *M. catarrhalis*). Therefore, our experiments indicate the potential of THz spectroscopy in combination with THz plasmonic antennas to detect bacteria selectively.

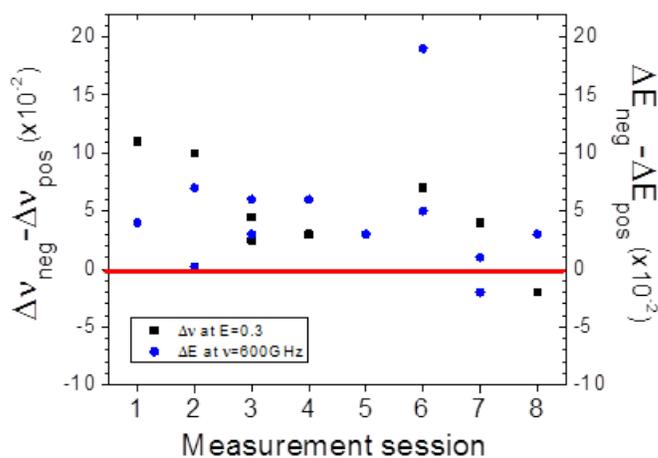

Figure 10 Difference between the behavior of the Gram negative compared to the Gram positive deposited plasmonic surfaces (high frequency antennas). The red line indicates the zero position.

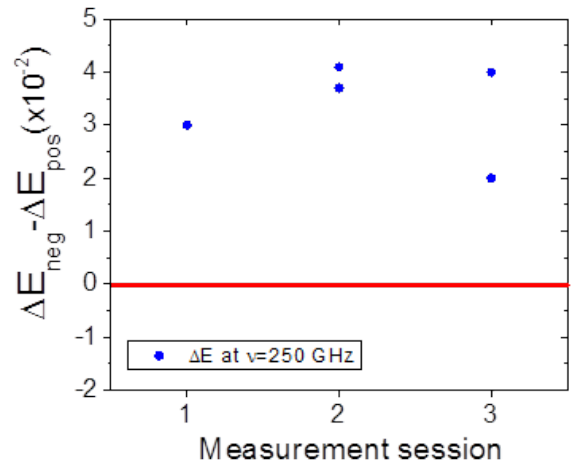

Figure 11 Difference between the behavior of the Gram negative compared to the Gram positive deposited plasmonic surfaces (low frequency antennas). The red line indicates the zero position.

It is important to note that we have performed the measurements with freshly grown and freshly prepared solutions of bacteria, in order to be sure that we are handling bacteria in a living state. Once the buffer solution with bacteria has been prepared, the solution was deposited and dried on the antenna samples and the THz transmission was measured within less than one hour. It is interesting to compare the difference in signal obtained by a freshly prepared bacterial layer with an 'old' bacterial layer. Therefore, we have deposited a bacterial layer on a low frequency THz antenna sample, measured the THz transmission following the standard procedure, and measured again the same sample after 24 hours at ambient air.

Figure 12 displays these THz measurements on antennas with a deposited layer of *E.coli*. The measurements on panel *a* have been performed within one hour after depositing the bacterial layer, i.e., the bacteria are still in the living state. Measurements in panel *b* have been performed more than 24 h (next day) after the deposition of the bacterial layer when the bacteria are no longer alive. The bacteria have most probably "died" by lysis, which implies a fragmentation of the cell wall and the escape of the intracellular fluid. We can see that the response of the antennas with a fresh layer of bacteria shows different spectral features than the response of the bare antennas, whereas the "dead" cells do not induce a modification of the antenna response (the spectral shift and features have disappeared). This result is interesting since it indicates that the THz plasmonic antenna response is sensitive to the living state of the bacteria. The difference seen in the THz signal can be explained by two facts: first, after lysis the bacteria lose their intracellular fluid, hence the THz absorption is decreased due to a lower water content of the layer; second, the structural change induced by the fragmentation of the cell wall influences the THz interaction with the constituents of the bacterial cell wall. The working principle of certain antibiotics is based on the lysis of the bacterial cells. Therefore, the sensitivity of the THz extinction of plasmonic antennas with bacteria to the structural shape of the cell wall can be a fast measure of the efficiency of given antibiotics.

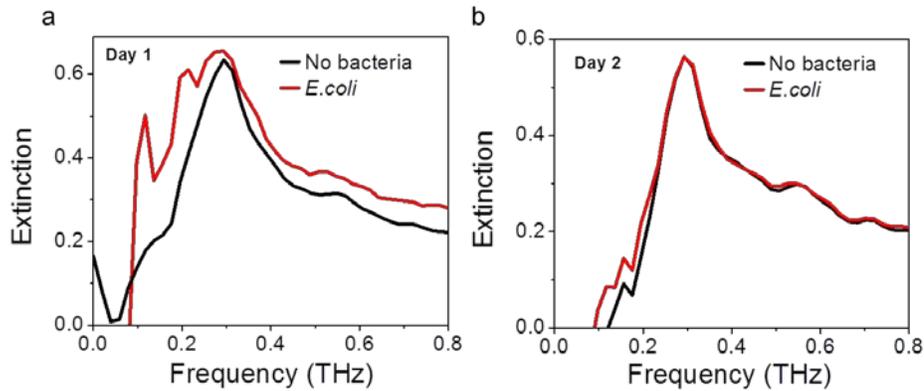

Figure 12 a) Extinction spectra of a low frequency operating THz antenna sample with freshly deposited bacterial layer; b) Extinction spectra of the same sample one day later.

**7. Conclusion**

We have investigated the interaction of bacterial layers on top of semiconductor plasmonic antennas. It was shown that the THz plasmonic antennas increase the sensitivity limit for thin films of biological matter and they allow the detection of bacteria down to a monolayer thickness. Furthermore, the presented method allows for the selective identification of the Gram type within the five different bacteria used in this study. This selectivity can be seen from experiments on double layers of bacteria, and from the analysis of multiple measurements on monolayers of bacteria. The origin of this difference in signal between Gram types is most likely due to structural difference in the cell wall or to a different content inside the bacteria. A more challenging situation would occur if the sample contains different types of bacteria. However, in situations where the bacterium to be identified forms the vast majority of the sample, such as body infections, the proposed technique can be of use. In these cases, statistical methods such as chemometrics or classification algorithms can be used to decide on the Gram type of an unknown sample.

In particular, in conjunction with a hand-held THz transceiver / receiver module, the use of THz plasmonic antennas has great potential for the definition of a label-free, low cost, and fast technique to detect and identify bacteria. The proposed technique is meant to complement other existing bacteria identification techniques by filling the gap of a rapid identification of infectious agents in point-of-care environments. The presented technique could also be useful in applications such as rapid antibiotic efficacy test.

**Acknowledgements**

We acknowledge L. Tripodi for fruitful discussions. This work was supported by the European Community's 7[th] Framework Programme under grant agreement n$^o$ FP7-224189 (ULTRA project) and is part of the research program of the "Stichting voor Fundamenteel Onderzoek der Materie (FOM)", which is financially supported by the "Nederlandse Organisatie voor Wetenschappelijk Onderzoek (NWO)". J.B. acknowledges the Swedish Research Council (Grant 621-2008-3562, 621-2011-4423) for financial support.